\documentclass[a4paper,11pt]{article}
\usepackage{soul}
\usepackage{jinstpub} 

\title{\boldmath PoWER: a new concept for DUNE Phase 2 FD PDS}

\author[a,1]{A. Steklain,\note{Corresponding author}}
\author[b]{E. Segreto}
\author[b]{A. Machado}
\author[a]{M. Adames}
\author[a]{L. Hirsch}
\author[c,d]{F. Di Capua}
\author[e]{N. Canci}
\author[b]{H. Frandini}

\affiliation[a]{Universidade Tecnológica Federal do Paraná, \\
Avenida Sete de Setembro 3165, Curitiba, Paran\'a, Brazil}
\affiliation[b]{Instituto de Fisica ``Gleb Wataghin'' Universidade Estadual de Campinas - UNICAMP,\\
Rua Sergio Buarque de Holanda, No 777, CEP 13083-859 Campinas, São Paulo, Brazil}
\affiliation[c]{Physics Department, Università degli Studi “Federico II” di Napoli, Napoli 80126, Italy}
\affiliation[d]{INFN Napoli, Napoli 80126, Italy}
\emailAdd{steklain@utfpr.edu.br}

\abstract{
We propose a novel concept for the future modules of the DUNE Phase 2 Far Detector Photodetection System, namely the Polymer Wavelength shifter and Enhanced Reflection - PoWER. In this concept, the field cage of the LArTPC is entirely covered with polymeric wavelength shifting foils (PolyEthylene Naphthalate – PEN) to convert the liquid argon scintillation light from VUV to visible, and an Enhanced Specular Reflector (ESR) is installed on the membrane aiming to increase the number of reflections and consequently the detection probability. In addition, we use Light Detection Units (LDUs), which are a combination of standard and VUV-sensitive SiPM that can be used as an active veto for events occurring outside the field cage. We present a preliminary study using a Monte Carlo simulation, including a Light Map for photons generated inside the field cage and a demonstration of the active veto.
}


\arxivnumber{1234.56789} 

\begin{document}
\maketitle
\flushbottom

\section{Introduction}
\label{sec:intro}

The Deep Underground Neutrino Experiment (DUNE) aims to probe CP violation in the neutrino sector and identify the neutrino mass hierarchy among other scientific objectives, that include new physics \cite{dunetdr4,dune}.
DUNE will employ a liquid argon time projection chamber (LArTPC) as the main technology for its detectors, consisting of near (ND) and far (FD) detectors 1,300 km from each other. The far detector will be composed of four LArTPC modules. The first two modules, currently under construction, will employ single-phase horizontal and vertical drift designs. For Phase 2, one additional vertical drift module is proposed together with a fourth module, which is referred to as a module of opportunity, and may be different. \cite{dunephase2}. 

The DUNE Photon Detection System (PDS) main capabilities lie in enhancing calorimetry and providing the event $t_0$ for drift time determination. Light detection might be used to obtain more information about the (neutrino) events inside the detector and improve physical searches: nucleon decay, supernova neutrino bursts, solar neutrinos, etc. To achieve the desired efficiency, the PDS  must meet two requirements: an average light yield larger than 20 PE/MeV, and a minimum light yield larger than 0.5 PE/MeV \cite{dunetdr4}.

The photodetection system for the second module of the DUNE far detector (FD2 PDS) employs X-ARAPUCAs installed on the membranes of the cryostat and in the cathode. Simulations indicate that this system provides an average light yield of 39 PE/MeV and a minimum light yield of 16 PE/MeV. For Phase 2, an improvement of this system is under development, called APEX. Simulations indicate that the APEX system can provide an average light yield of 180 PE/MeV, with a minimum of 109 PE/MeV \cite{dunephase2}. The APEX system uses X-ARAPUCAs \cite{Machado2016,Machado2018}, a photon collector that uses a combination of wavelength shifter deposited in the acceptance window and a dichroic filter. Despite its efficiency, producing it on a large scale is very complex.

We propose a novel concept for the Phase 2 FD PDS, the Polymer Wavelength shifter and Enhanced Reflection - PoWER, which has many interesting features for the experiment. First, it has a very simple design, with the SiPMs lying outside the field cage (and the electric field inside of it), which simplifies their operation and maintenance. Second, it is cost-effective, being composed of low-cost components that require little research and development. Finally, our preliminary study using a standalone Geant4 simulation indicates that this system provides the necessary light yield necessary for DUNE PDS scientific goals. In the next sections, we present PoWER and the preliminary results of the Geant4 simulation.

\section{PoWER Photodetection System}

Liquid argon produces abundant ionization and scintillation light.
In a TPC operated at 500 V/cm, about half of the energy deposited by ionizing particles goes into photons. Nevertheless, the scintillation light is produced with a wavelength of 128 nm, in the vacuum ultraviolet region (VUV). The light is severely scattered by liquid argon in this wavelength interval, jeopardizing signal uniformity.

To address this issue, it is possible to improve the uniformity of light collection by doping the liquid argon with xenon. At a concentration of 10 ppm of Xe the slow scintillation component is almost completely shifted to 173 nm, where the Rayleigh scattering length is four times larger than at 128 nm. In order to increase the collected light, detectors like X-ARAPUCA rely on wavelength-shifting (WLS) materials, in particular paraTherPhenyl (PTP), to convert argon scintillation light produced at VUV to visible wavelengths, enabling efficient detection. PTP is applied on a substrate by vacuum deposition, making it difficult to use in large areas. Poly(ethylene 2,6-naphthalate) (PEN) is being developed as a replacement for PTP. PEN has been shown to have 30\% to 75\% of the efficiency of the most commonly used shifter in combination with LAr, TetraPhenylButadiene (TPB) for shifting VUV photons to visible wavelengths \cite{Kuzniak2019}. In addition, it is a thermoplastic polyester that can be easily drawn into foils or extrusion molded.

One key advantage of the PoWER system is its simple design. In this concept, the field cage is entirely covered with polymeric foils made of a 0.1 mm layer of PolyEthylene Naphthalate (PEN) and a 1 mm layer of acrylic. Large plastic panels lined with Enhanced Specular Reflector (ESR) material—offering 95\% reflectivity in the visible spectrum—are installed on the membrane to enhance detection probability. The cathode is partially covered with PEN foils and ESR, which include perforations to enable LAr circulation.

\begin{figure}[htbp]
\centering
\includegraphics[width=.8\textwidth]{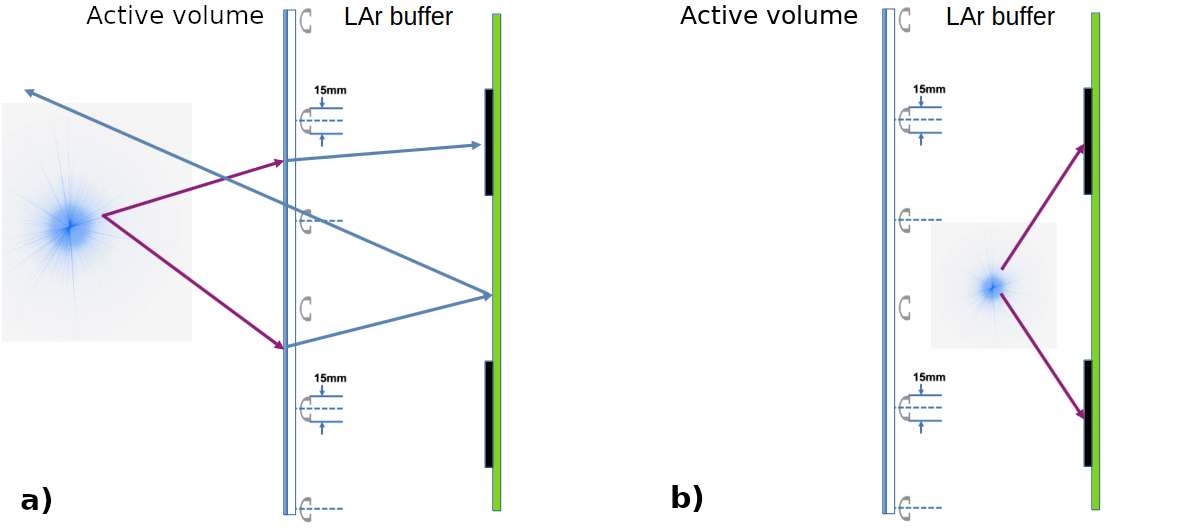}
\caption{PoWER photodetection system. a) If the event takes place in the active volume, the scintillation light is converted by the PEN and is detected by the LDUs or reflected back to the active volume. b) Active veto system. If the event takes place in the buffer, the scintillation light is blocked by the acrylic and is not converted, eventually detected by the VUV-sensitive component of the LDUs only. \label{fig:power}}
\end{figure}

The principle is illustrated in figure~\ref{fig:power}a). If the event takes place in the active volume (inside the field cage), the scintillation light that travels in the direction of the field cage or cathode is converted by the PEN layer from VUV to visible light. After conversion, the light may be detected or reflected by the ESR back to the active volume. As the Rayleigh scattering length for visible light is large, the light can cross the volume and be detected on the other sides.

The detection is carried out by 1,872 light detection units (LDUs) installed on the cryostat membrane outside the field cage. Each LDU comprises an array of SiPMs, including two different components that are sensitive to visible and VUV light. This scheme allows for an active veto for events in the liquid argon (LAr) buffer between the field cage and the membrane. If the event takes place in the buffer zone, as shown in figure~\ref{fig:power}b), the VUV light produced is detected only by some of the SiPMs on the LDUs, providing an active veto system for events happening outside of the field cage. In this case, the acrylic does not allow the VUV light to enter the active volume and be converted to visible light. The scintillation light is detected by the VUV-sensitive SiPMs only. The veto is then performed if the signal from these SiPMs is much larger than the visible-only sensitive SiPMs.


\section{Geant4 Simulation}

\begin{figure}[htbp]
    \centering
    \includegraphics[width=0.9\linewidth]{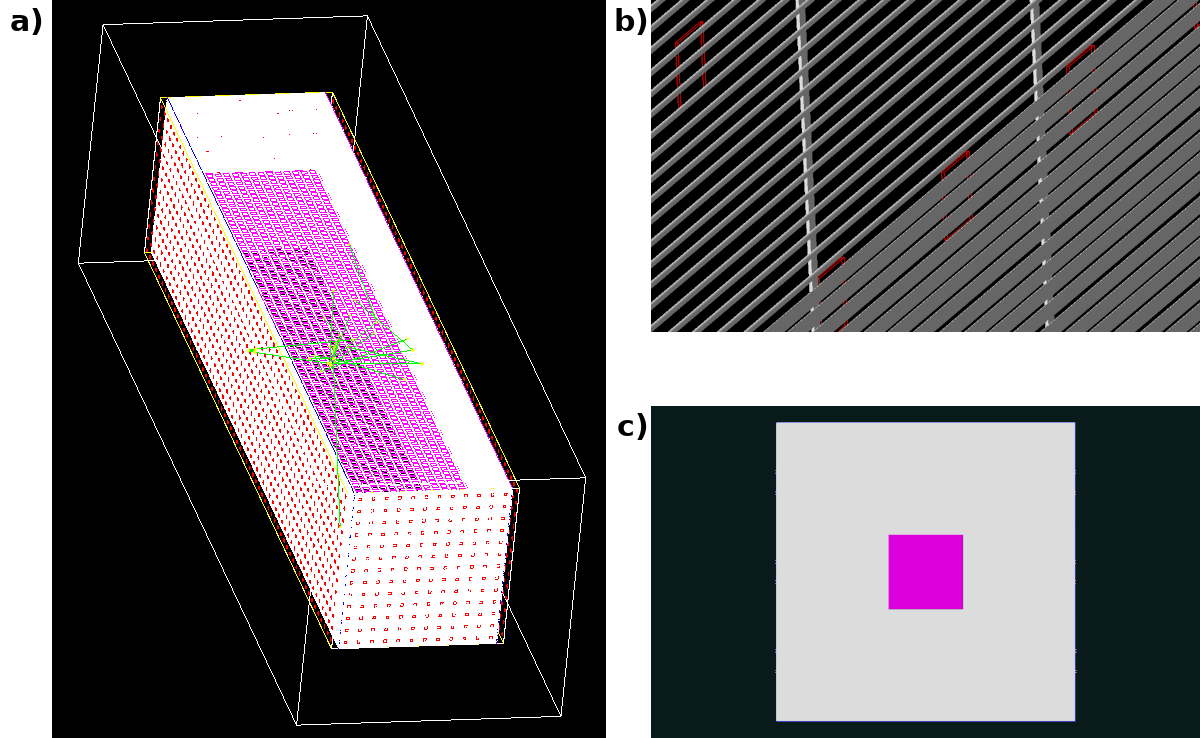}
    \caption{Geant4 simulation for PoWER photodetection system. a) Overview of the simulation. b) Visualization of the two transparency regions of the field cage in the simulation. c) Geometry of the LDU. }
    \label{fig:simpower}
\end{figure}

We use a standalone Geant4 simulation to study the light yield of the PoWER system. The overview of the simulation can be found in figure~\ref{fig:simpower}a). In this simulation, the cryostat is represented by a 13 m $\times$ 13 m $\times$ 60 m box
filled with a mixture of liquid argon and xenon, the last one in the proportion of one part per million. The walls of the cryostat are covered by ESR with 98\% reflectivity. The bottom and the top of the box are covered by a material with 30\% reflectivity, which represents the average reflectivity of the charge read-out panels. 

The field cage is a metallic mesh that follows the specifications from DUNE. The spacing between profiles allows for some transparency in the field cage. There are two distinct regions of 35\% and 75\% profile transparency, as shown in figure~\ref{fig:simpower}b). In the center, there is the cathode, represented by a grid with square cells with sides measuring 60 cm. In front of the four sides of the field cage, there are walls composed of a 0.1 mm thick layer of PEN and a 1 mm thick layer of acrylic. The material's optical characteristics are the same as used in Legend-200 simulations \cite{Manzanillas2022}. The cathode is also partially covered by an ESR and PEN layer, with 42cm side square holes. The LDUs are modeled using the specifications of two distinct Hammamatsu SiPMs. For the visible, we use the specifications of Model S14160-6050HS, and for VUV, we use Model VUV4 S13370. The geometry of the LDUs is shown in figure~\ref{fig:simpower}c). Each LDU has a total area of 20 cm $\times$ 20 cm, with the center area of 5 cm $\times$ 5 cm occupied by VUV-sensitive SiPMs and the remaining area occupied by visible-only sensitive SiPMs. The LDUs cover a total area of 75 m$^2$ in the cryostat.

\begin{figure}[htbp]
    \centering
    \includegraphics[width=0.6\linewidth]{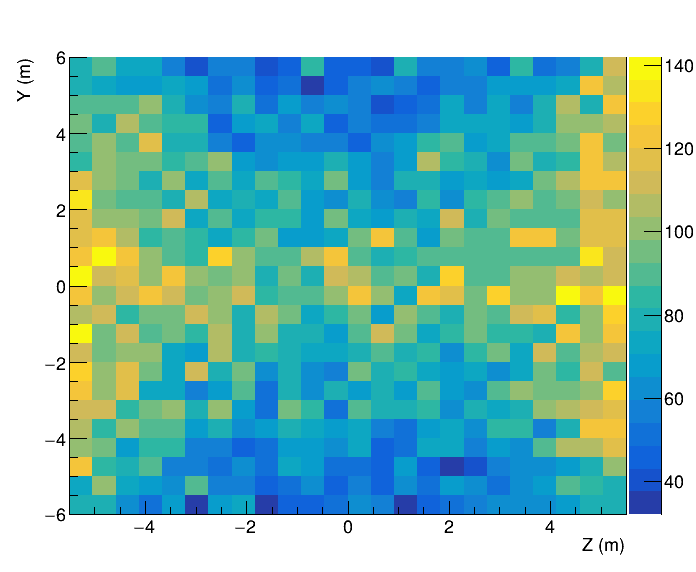}
    \caption{Light map for the active region of the cryostat. }
    \label{fig:lightmap}
\end{figure}

The light map is generated using a total of 14.4 million photons, in the center of the cryostat, resulting in approximately 25,000 photons/MeV per voxel of dimensions $0.5\text{m} \times 0.5\text{m} \times 0.5\text{m}$. In the simulation, each time a photon is detected by an LDU the coordinates of its origin are registered. We consider the contribution of both types of SiPMs. The result is shown in the figure~\ref{fig:lightmap}. The simulation indicates an average light yield of 90 PE/MeV, and a minimum light yield of 35 PE/MeV, an improvement in comparison to FD2 PDS.

\begin{figure}[htbp]
    \centering
    \includegraphics[width=0.6\linewidth]{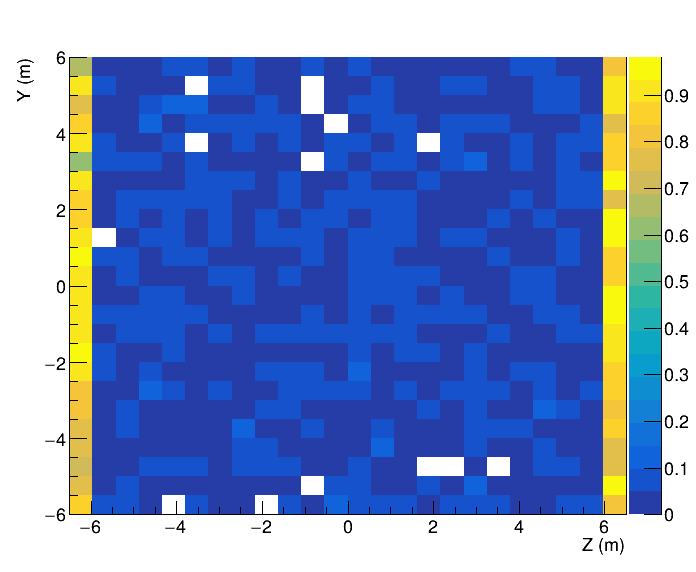}
    \caption{PoWER veto system. Events inside the active volume have $r_{VUV} < 0.3$, as events in the buffer zone have $r_{VUV} > 0.6$. White squares correspond to $r_{VUV} = 0$.}
    \label{fig:veto}
\end{figure}

The PoWER veto system is obtained by considering the distinct contributions from VUV and visible-only sensitive SiPMs. Considering the relative contribution of VUV given by
\begin{equation}
    r_{VUV} = \frac{LY_{VUV}}{LY_{VUV}+LY_{Visible}},
\end{equation}
we expect that events that take place inside the active volume should have $LY_{VUV}\ll LY_{Visible}$, as the VUV-sensitive SiPMs have a lower sensitivity to visible light, and the relative area covered by these SiPMs is 1/15 of the area covered by visible-only sensitive SiPMs. It should result in a low value of $r_{VUV}$. On the other hand, if the event takes place in the buffer zone, we should have $LY_{VUV}\gg LY_{Visible}$, as the visible-only sensitive SiPMs are unable to detect VUV light. In this case, we should have a high value of $r_{VUV}$. The situation is illustrated in figure~\ref{fig:veto}, showing that $r_{VUV}$ value for events inside and outside the field cage. We see a clear distinction between the two types of events, showing how the veto system may work.

The first results from the simulation indicate that PoWER may serve as a photodetection system for DUNE Phase 2. The light yield of this system is already superior to that of DUNE FD3 PDS, and it uses a much simpler setup. There is still much room for improvement in this setup, such as optimizing the LDU covering and improving PEN efficiency. A small-scale prototype will be proposed to demonstrate the veto for the next steps. Also, new simulations will study the improvement of the average light yield.

\acknowledgments

The authors gratefully acknowledge the financial support provided by the National Council for Scientific and Technological Development (CNPq) through a Scientific Initiation Scholarship (PIBIC). This study was financed in part by the Fundação Araucária and SETI through grant PDI 346/2024. It was also partly funded by São Paulo Research Foundation (FAPESP). Their support was essential for the development of this research.


\bibliographystyle{JHEP}
\bibliography{biblio.bib}


\end{document}